\documentclass[
    ,final            
]
  {aipproc}
\usepackage{bm}
\layoutstyle{6x9}

\newcommand{\tr}{{\textrm {tr}}}

\newcommand{\Tr}{{\textrm {Tr}}}

\newcommand{\SU}{{\textrm{SU}}}

\begin{document}

\title{The Quantum and Local Polyakov loop in Chiral Quark Models at
Finite Temperature~\thanks{Supported by funds provided by the Spanish
DGI and FEDER founds with Grant No. FIS2005-00810, Junta de
Andaluc\'{\i}a Grant No. FM-225, and EURIDICE Grant
No. HPRN-CT-2003-00311.}}

\author{\underline{E. Meg\'{\i}as}~\thanks{Speaker at Quark Confinement and
  the Hadron Spectrum VII, Ponta Delgada, Portugal, 2-7 IX, 2006.}, 
E. Ruiz Arriola and L.L. Salcedo}
{address={Departamento de F{\'{\i}}sica At\'omica, Molecular y Nuclear,  
Universidad de Granada. \\ 18071-Granada (Spain)}
}
\begin{abstract}
  We describe results for the confinement-deconfinement phase
  transition as predicted by the Nambu--Jona-Lasinio model where the
  local and quantum Polyakov loop is coupled to the constituent quarks
  in a minimal way (PNJL). We observe that the leading correlation of
  two Polyakov loops describes the chiral transition accurately. The
  effects of the current quark mass on the transition are also
  analysed.
\end{abstract}


\maketitle




The simultaneous ocurrence of two phase transitions where chiral
symmetry is restored and hadrons become deconfined seems a unique and
misterious feature of QCD matter at finite temperature (for an early
review see e.g. ~\cite{Karsch:1998hr}). Besides direct QCD lattice
simulations, there exist theoretical constraints below and above the
deconfinement phase transition. At low temperatures the leading
thermal excitations correspond to a gas of weakly interacting
pions~\cite{Gerber:1988tt}. Moreover, in the large $N_c$ limit with
the temperature $T$ kept fixed, if a chiral phase transition takes
place it should be first order~\cite{Gocksch:1982en}. The use of
resonance hadron Lagrangians implies that thermal corrections are
$1/N_c$ suppressed~\cite{Toublan:2004ks}.  
At high temperatures one has a weakly interacting
quark-gluon plasma (for a review see e.g. \cite{Kraemmer:2003gd}).
However, the previous powerful constraints assume from the start a
given phase and do not provide a clue 
on how chiral symmetry restoration and deconfinement are intertwined.

The coupling of relevant order parameters such as the quark condensate
for chiral symmetry breaking and the Polyakov loop for deconfinement
at finite temperature can be made explicit in Polyakov chiral quark
models, an amalgamate of colour and flavour degrees of freedom where
the simultaneous chiral-deconfinement crossover can be quantitatively
studied with an acceptable phenomenological success
\cite{Meisinger:1995ih,Meisinger:2003uf,Fukushima:2003fw,Megias:2004kc,
Megias:2004hj,Ratti:2005jh,Ghosh:2006qh,Megias:2006bn,Hansen:2006ee,Ratti:2006wg}.
Actually, in our recent work~\cite{Megias:2004hj} we have shown why
and how Polyakov loops must be coupled to Chiral Quark Models (CQM) to
comply with large gauge invariance at finite temperature and how the
quantum and local nature of the Polyakov loop generates a rather sharp
crossover at about the observed critical temperature, although
uncertainties are expected. More specifically, ChPT and large $N_c$
constraints are naturally accomodated~\cite{Megias:2006bn} within
those models, hence solving a long standing puzzle which was ignored
for a long time; traditional CQM did produce a chiral phase transition
while violating those restrictions. An immediate consequence of this
new coupling is an upward shift of the critical temperature referred
to as {\it Polyakov cooling} in Ref.~\cite{Megias:2004hj}.


In this work we will deal with the PNJL model for definiteness.  After
bosonization the NJL Lagrangian reads
\begin{equation}
\Gamma_Q [S,P] =- {\rm Tr} \log \left( i\slash\!\!\!\!\!{\partial} 
+ {\hat M_0} +  S + i\gamma^5 P  \right)  
+ {1\over 4G_S } \int d^4 x   \,{\rm tr}_f \left( S^2
+ P^2 \right) \,.
\label{eq:eff_ac_njl} 
\end{equation} 
We use ${\rm Tr}$ for the full functional trace, $\tr_f $ for the
trace in flavour space, and $\tr_c $ for the trace in colour
space. The UV divergencies in Eq.~(\ref{eq:eff_ac_njl}) from the Dirac
determinant only affect in practice the zero temperature
contributions~\cite{Christov:1991se}. In the Polyakov gauge $\Omega=
e^{i A_4/T}$, where $A_4$ is time independent and diagonal, and the
minimal coupling is made $ \partial_4 \to \partial_4 - i A_4 $.
Integrating further over the $A_4$ gluon field in a gauge invariant
manner~\cite{Reinhardt:1996fs} yields a generic partition function
of the form
\begin{eqnarray}
Z = \int DS \, DP \, D\Omega \; e^{- \Gamma_G [\Omega]} e^{- \Gamma_Q [ S,P;
\Omega ]} \,,
\label{eq:Z_pnjl} 
\end{eqnarray}
where $ D \Omega $ is the Haar measure of the $\SU(N_c)$ colour group,
$\Gamma_G $ is the effective gluon action and $\Gamma_Q$ stands for
the quark effective action. In general $\Omega (x) $ is a local and
quantum variable since the gluon field itself depends on the point. As
argued in ~\cite{Megias:2004hj} mean field
approximations~\cite{Meisinger:2003uf,Fukushima:2003fw,Ratti:2005jh,Ghosh:2006qh}
generate a spurious gauge orbit dependence and a possibly complex
$\Omega$ (violating colour charge conjugation) and they necessarily imply
a non-vanishing value of the Polyakov loop in the adjoint
representation, in contradiction with lattice results~\cite{Dumitru:2003hp}.


For a constant value of the Polyakov loop, $\Omega$, and the scalar
field $S=M$ (which we identify with the constituent quark mass) the
quark effective action is given by 
\begin{eqnarray}
\frac{ T }{V} \Gamma_Q ( M, \Omega,T )&=& 
\frac{1}{4 G_S} \Tr_f(M - \hat{M}_0)^2  - 2 N_f 
\int \frac{d^3 k}{(2\pi)^3} \Bigg[ N_c \epsilon_k \nonumber \\
&& +  T \, {\rm Tr}_c \left\{ \log\left( 1 +
e^{-\epsilon_k/T} \Omega \right) + \log\left( 1 + e^{-\epsilon_k/T}
\Omega^\dagger \right) \right\} \Bigg]  \,,
\label{eq:gammaQ1}
\end{eqnarray} 
where we have only retained the vacuum contribution, so there is no
contribution of meson fields $(S,P)$ (see \cite{Megias:2006bn} for a
chiral expansion up to ${\cal O}(p^4)$), $V$ is three dimensional
volume and $\epsilon_k = +\sqrt{k^2 + M^2 } $ is the energy of a
constituent quark with mass $M$. We define the Polyakov-loop averaged
action
\begin{eqnarray}
e^{-\Gamma_Q (M,T)} = \int d \Omega \, e^{-\Gamma_G[\Omega]}
e^{-\Gamma_Q (M,\Omega,T)} \,.
\end{eqnarray} 
The value of $M$ is determined by minimization of $\Gamma_Q(M,T)$ with
respect to $M$, $ \partial \Gamma_Q (M,T) / \partial M=0  $, which
corresponds to computing the integration in $DS DP$ at the mean field
level and determines $M$ at a given temperature $T$, denoted as $M^* =
M(T) $. In addition, the relation between the (single flavour) chiral
quark condensate, $\langle \bar q q \rangle$, and the constituent
quark mass, reads 
\begin{eqnarray}
2 G_S N_f \langle \bar q q \rangle^* = - (M^* - m_q) \,.
\label{eq:gap*}
\end{eqnarray} 
Any observable is obtained by using $M^*$ and averaging over
$\Omega$. The integral in $d\Omega$ in the case $N_c=3$ and in the
Polyakov gauge was computed numerically in Ref. \cite{Megias:2004hj}.
Here we show a much simpler method which is based on evaluating the
integral analytically for any $N_c$ in the low temperature limit and
corresponds to take $\Omega$ small. (To see this use the formula $
\int d^3 k \, e^{-\epsilon_k /T} = 4 \pi M^2 T K_2 \left( \frac{M}{T}
\right) \,, $ where $K_2(x)$ is the modified Bessel function and $ K_2
(x) \sim \sqrt{\pi/2\, x} \, e^{-x} $ for $x \to \infty $). From
Eq.~(\ref{eq:gammaQ1}) we get
\begin{eqnarray}
 \Gamma_Q (M,\Omega,T) &=& 
 \Gamma_Q (M,0) + 2 N_f \sum_{n=1}^\infty 
\frac{(-1)^n}{n} \int \frac{d^3x\, d^3 k}{(2\pi)^3}
e^{-n \epsilon_k /T}  {\rm Tr}_c [ \Omega^n+ 
\Omega^\dagger{}^n ]  \,.
\end{eqnarray} 
Expanding the exponent in Eq.~(\ref{eq:Z_pnjl}) one obtains a power series in
terms of $\Omega $ and $ \Omega^\dagger $. The simplest
correlation of two Polyakov loops is taken to be~\cite{Megias:2004hj}
\begin{eqnarray}
\int d\Omega \,\Tr_c \Omega(\vec{x}) \Tr_c \Omega^\dagger (\vec{y}) 
= e^{-\sigma |\vec{x}-\vec{y}|/T} \,,
\label{eq:corLL}
\end{eqnarray}
with $\sigma$ the string tension, and yields the leading thermal
contribution to the effective action
\begin{eqnarray}
\frac{T}{V}\Gamma_Q (M,T) =\frac{T}{V}\Gamma_Q (M,0)- T V_\sigma \left( 2 N_f
\int \frac{d^3 k}{(2\pi)^3} e^{-\epsilon_k /T} \right)^2 + \cdots  \,,
\label{eq:eff_ac_om} 
\end{eqnarray} 
where $V_\sigma= 8\pi T^3/\sigma^3$ is the equivalent confinement
correlation volume. As we see, the effect of quantum corrections on
the Polyakov loop lowers the vacuum energy as it should. Moreover,
they are $1/N_c$ suppressed, as one would expect in Chiral
Perturbation Theory or a resonance gas model but unlike traditional
chiral quark models without Polyakov loop.

\begin{figure}[ttt]
\begin{minipage}[t]{7.3cm}
\includegraphics[height=4.5cm,width=6.5cm]{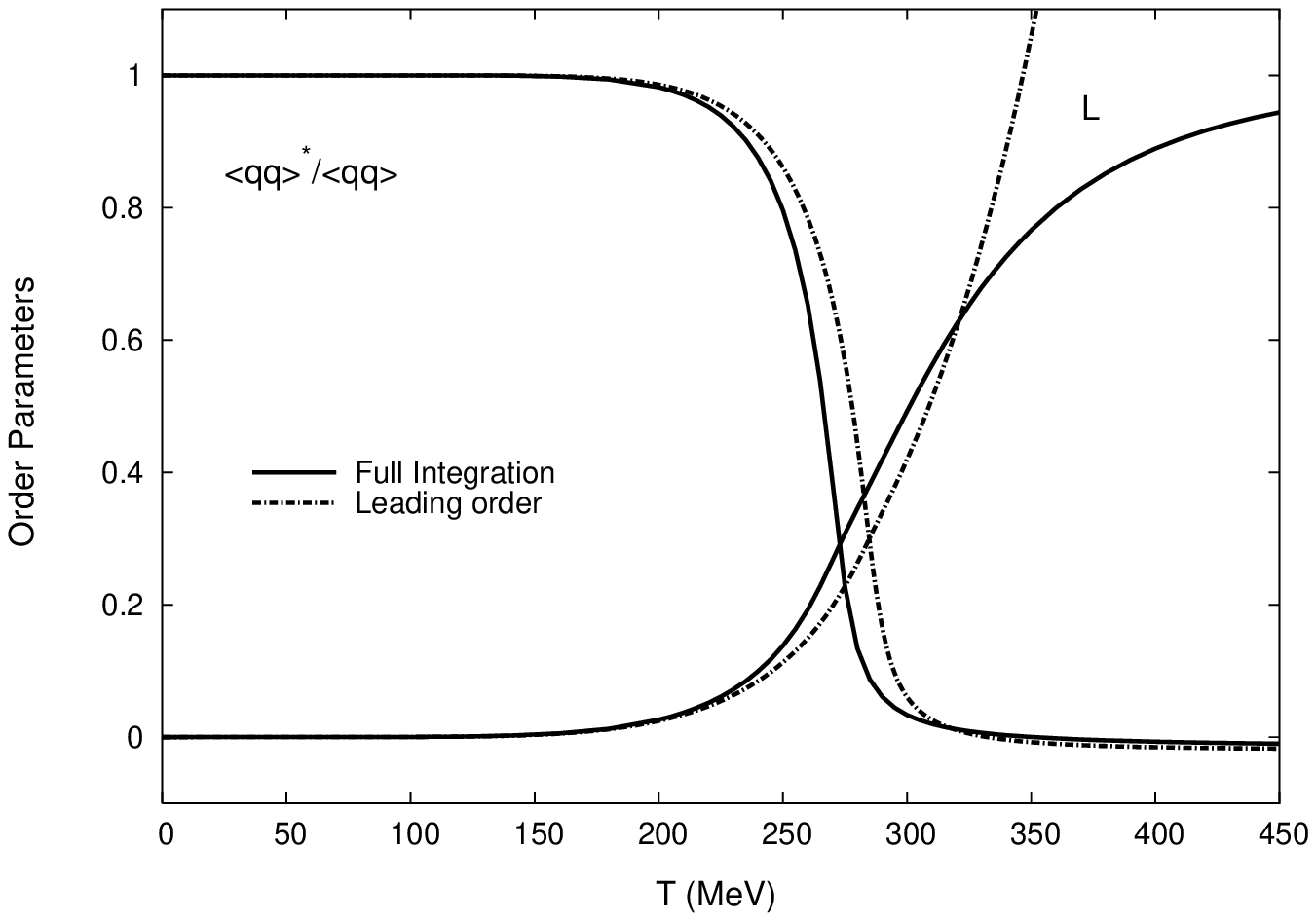}
\end{minipage}
\hspace*{0.3cm}
\begin{minipage}[t]{7.3cm}
\includegraphics[height=4.5cm,width=6.5cm]{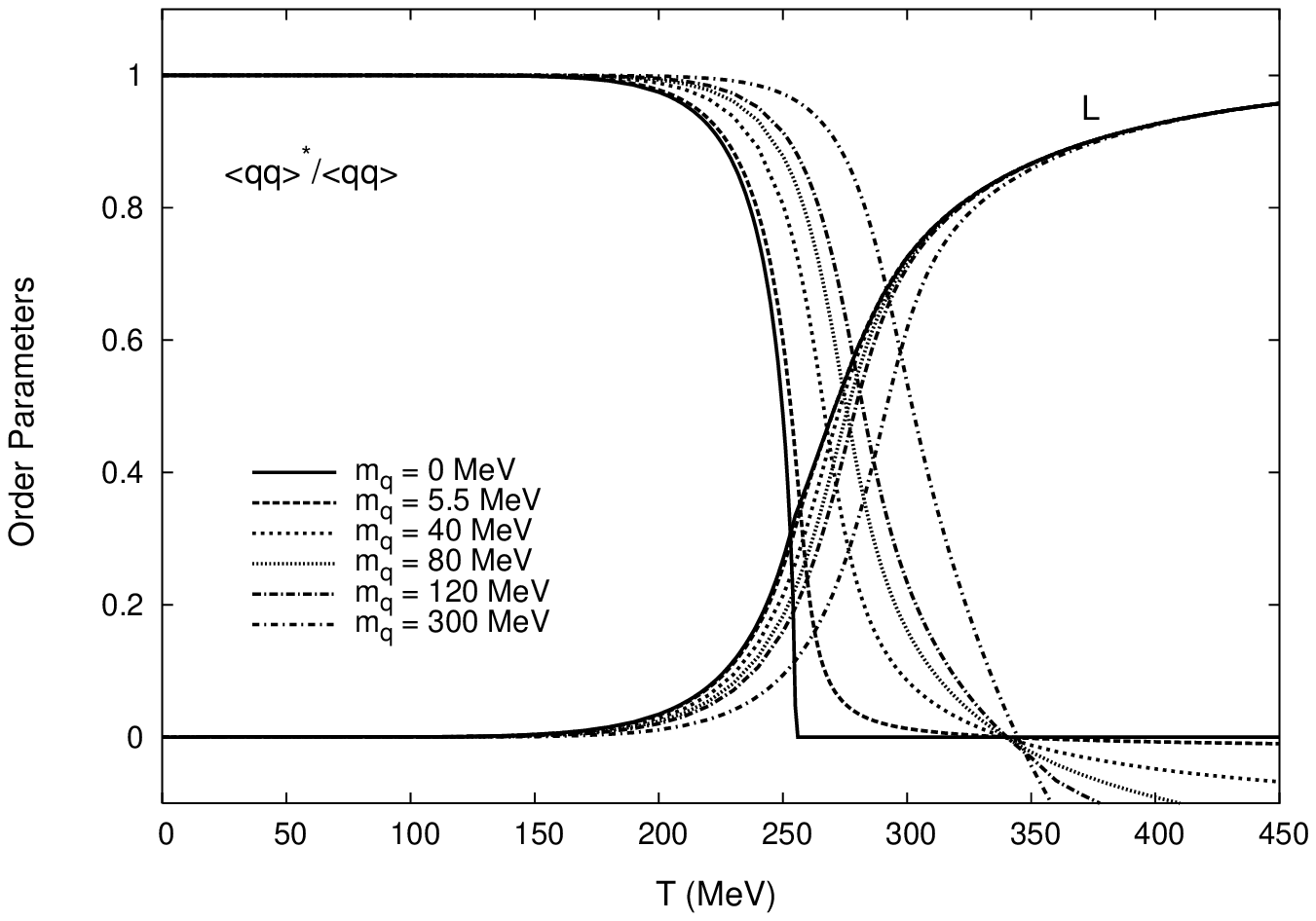}
\end{minipage}
\caption{Chiral condensate $\langle \bar q q \rangle^* $ and 
  Polyakov loop $L = \left\langle \tr_c \Omega\right\rangle / N_c$. 
Left: Leading Polyakov loop correlation approximation (see
  Eqs.~(\ref{eq:eff_ac_om}) and (\ref{eq:L_NJL_lowT})). We take the
  2-flavor PNJL model, and $\sqrt{\sigma} = 425 \,{\rm MeV}$, $f_\pi = 93
  \,{\rm MeV}$, $M=300 \,{\rm MeV}$, $m_u = m_d \equiv m_q=5.5 \,{\rm MeV}$.
  Right:
 Current quark mass dependence. 
}
\label{fig:qq_app}
\end{figure}

Minimizing with respect to the mass we get the effective temperature
dependent mass $M^*$ and from Eq.~(\ref{eq:gap*}) the corresponding
$\langle \bar q q \rangle^*$ condensate can be evaluated. The
approximate result is presented and compared to the full
result~\cite{Megias:2004hj} in Fig.~\ref{fig:qq_app} and, as we can see
the approximation is quite efficient and very easy to implement in
standard chiral quark models. For the Polyakov loop expectation value
similar manipulations hold, yielding the leading order contribution
\begin{eqnarray} 
L = \left\langle\frac{1}{N_c} \tr_c \Omega\right\rangle=
\frac{N_f}{N_c} V_\sigma  \frac{M^2 T }{\pi^2} K_2 (M /T)
+ \cdots &\sim & \frac{N_f}{N_c} \frac{V_\sigma}{T} 
\sqrt{\frac{M^3 T^5}{2\pi^3}} e^{-M/T} \,.
\label{eq:L_NJL_lowT}
\end{eqnarray}
The full result and the approximated formula are compared in
Fig.~\ref{fig:qq_app}. In this case the agreement is only up to
temperatures about $0.75 T_D$. 

\begin{figure}[ttt]
\begin{minipage}[t]{7.3cm}
\includegraphics[height=4.5cm,width=6.5cm]{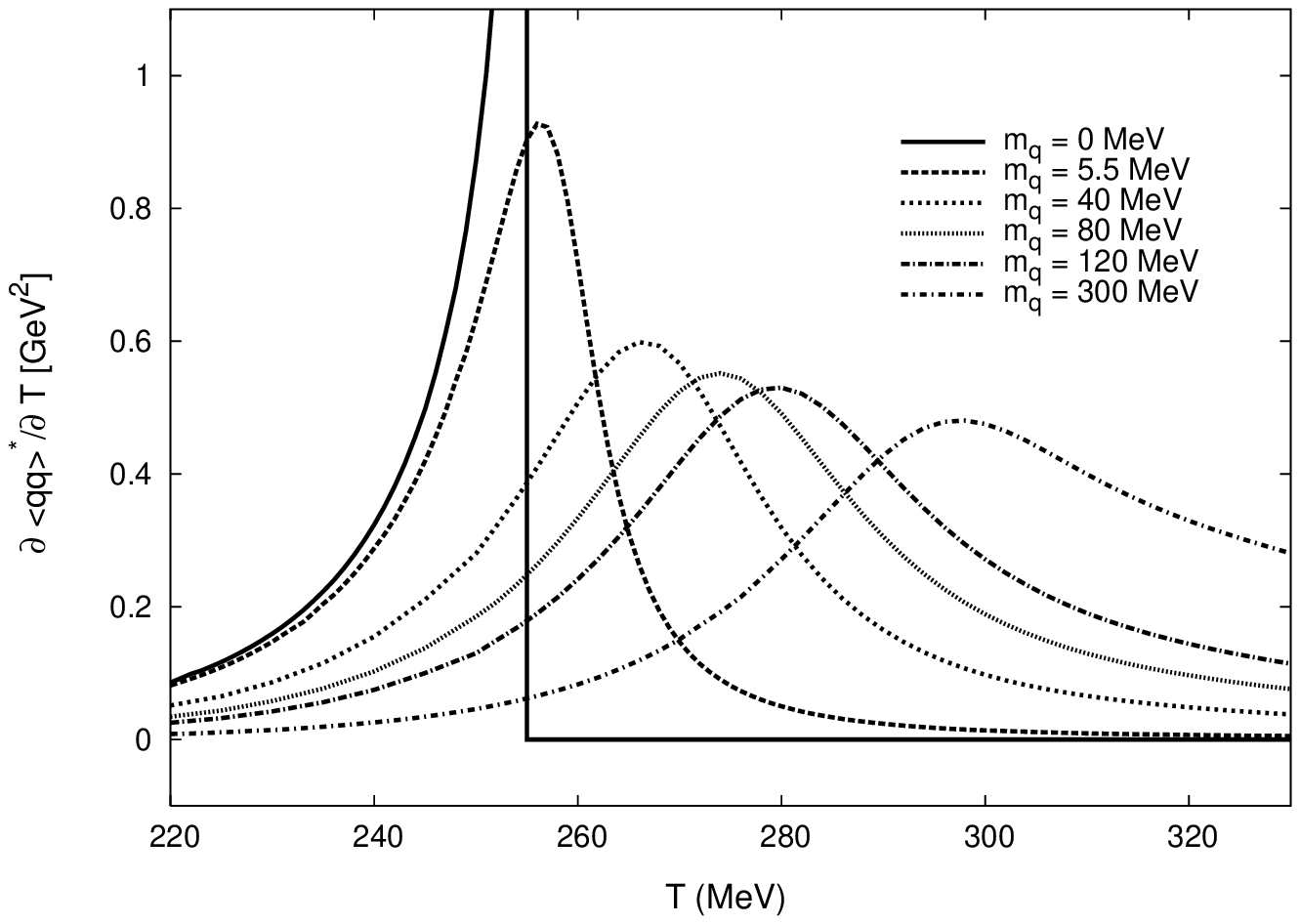}
\end{minipage}
\hspace*{0.3cm}
\begin{minipage}[t]{7.3cm}
\includegraphics[height=4.5cm,width=6.5cm]{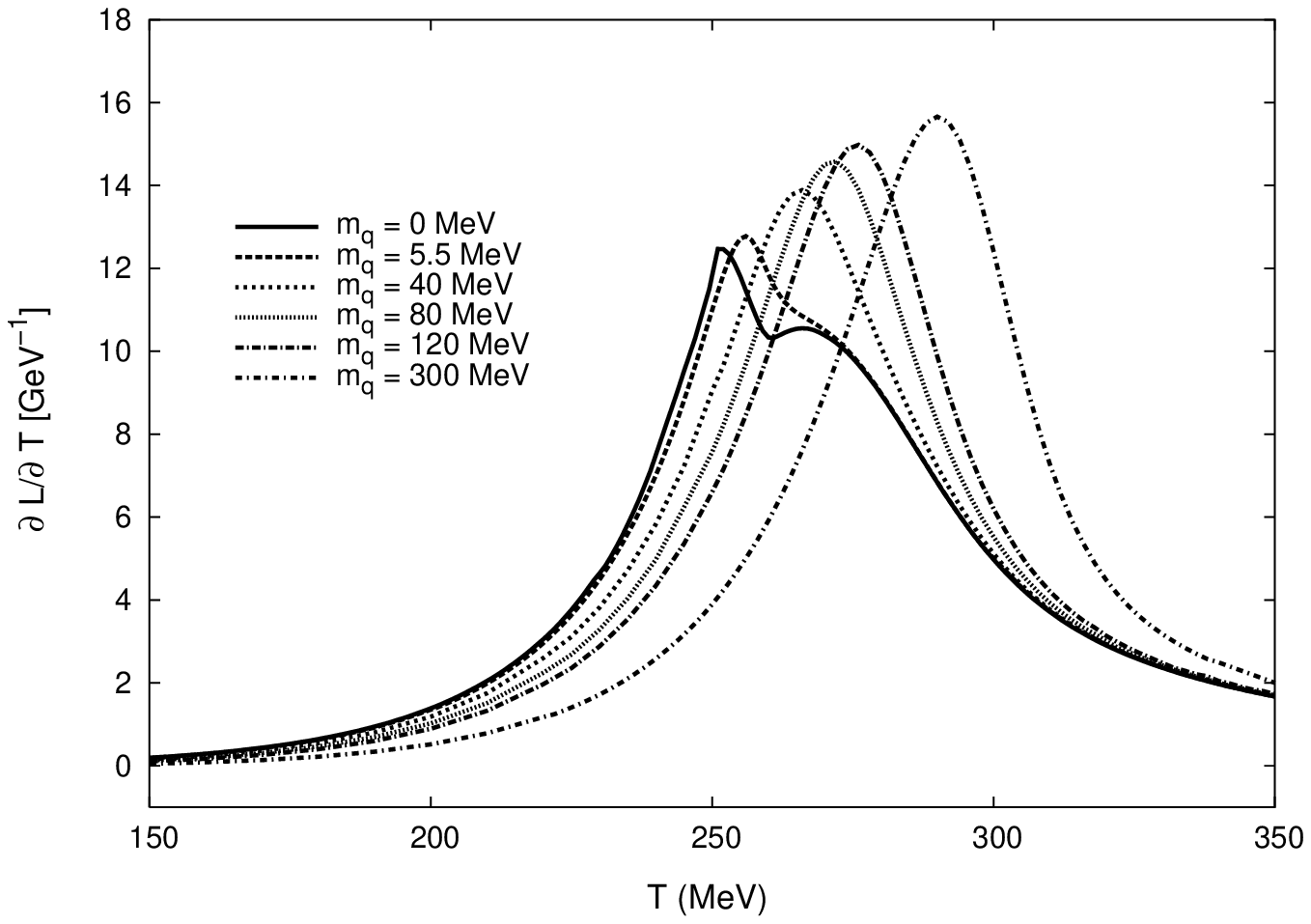}
\end{minipage}
\caption{Temperature dependence of $\partial L/\partial T$ (left) and 
$\partial \langle \bar q q\rangle^*/\partial T$ (right) for several values of
the current quark mass. The transition temperatures are $T_\chi = T_D =  255(1),
\;256(1),\;
266(1) \; {\rm MeV}$ for $m_q= 0,\; 5.5,\; 40 \; {\rm MeV}$ respectively, and $T_\chi
= 297(1) \; {\rm MeV}$, $T_D=290(1) \; {\rm MeV}$ for $m_q=300\;{\rm MeV}$.}
\label{fig:Omega_mq}
\end{figure}

The analysis above is done with physical current quark masses, $m_q=5.5\, {\rm
  MeV}$ where one obtains $T_\chi = T_D = 256(1) \,{\rm MeV}$. This remarkable
coincidence between transitions is not accidental nor depends on the
particular choice of $m_q$ as can be seen in Fig.~\ref{fig:qq_app}, where we
show the temperature dependence of $\langle \bar q q\rangle^* $ and $L$ for
$m_q=0,\; 5.5, \; 40, \; 80, \; 120 $ and $300\; {\rm MeV}$. The corresponding
susceptibilities are displayed in Fig.~\ref{fig:Omega_mq}.

  


\end{document}